# Status Quo and Problems of Requirements Engineering for Machine Learning: Results from an International Survey

Antonio Pedro Santos Alves[1], Marcos Kalinowski[1], Görkem Giray[2], Daniel Mendez[3], Niklas Lavesson[3], Kelly Azevedo[1], Hugo Villamizar[1], Tatiana Escovedo[1], Helio Lopes[1], Stefan Biffl[4], Jürgen Musil[4], Michael Felderer[5,6], Stefan Wagner[7], Teresa Baldassarre[8], and Tony Gorschek[3]

[1] Pontifical Catholic University of Rio de Janeiro (PUC-Rio), Brazil
[2] Independent Researcher, Turkey
[3] Blekinge Institute of Technology (BTH), Sweden
[4] Vienna University of Technology (TU Wien), Áustria
[5] German Aerospace Center (DLR), Germany
[6] University of Cologne, Germany
[7] University of Stuttgart, Germany
[8] University of Bari, Italy

**Abstract.** Systems that use Machine Learning (ML) have become commonplace for companies that want to improve their products and processes. Literature suggests that Requirements Engineering (RE) can help address many problems when engineering ML-enabled systems. However, the state of empirical evidence on how RE is applied in practice in the context of ML-enabled systems is mainly dominated by isolated case studies with limited generalizability. We conducted an international survey to gather practitioner insights into the status quo and problems of RE in ML-enabled systems. We gathered 188 complete responses from 25 countries. We conducted quantitative statistical analyses on contemporary practices using bootstrapping with confidence intervals and qualitative analyses on the reported problems involving open and axial coding procedures. We found significant differences in RE practices within ML projects. For instance, (i) RE-related activities are mostly conducted by project leaders and data scientists, (ii) the prevalent requirements documentation format concerns interactive Notebooks, (iii) the main focus of non-functional requirements includes data quality, model reliability, and model explainability, and (iv) main challenges include managing customer expectations and aligning requirements with data. The qualitative analyses revealed that practitioners face problems related to lack of business domain understanding, unclear goals and requirements, low customer engagement, and communication issues. These results help to provide a better understanding of the adopted practices and of which problems exist in practical environments. We put forward the need to adapt further and disseminate RE-related practices for engineering ML-enabled systems.

## 1  Introduction

Companies from different sectors are increasingly incorporating Machine Learning (ML) components into their software systems. We refer to these software systems, where an ML component is part of a larger system, as ML-enabled systems. The shift from engineering conventional software systems to ML-enabled systems comes with challenges related to the idiosyncrasies of such systems, such as addressing additional qualities properties (*e.g.,* fairness and explainability), dealing with a high degree of iterative experimentation, and facing unrealistic assumptions [21, 25]. Furthermore, the non-deterministic nature of ML-enabled systems poses challenges from the viewpoint of software engineering [7].

Literature suggests that Requirements Engineering (RE) can help to address problems related to engineering ML-enabled systems [1,25,28]. However, research on this intersection mainly focuses on using ML techniques to support RE activities rather than exploring how RE can improve the development of ML-enabled systems [4]. The state of empirical evidence on how RE is applied in practice in the context of ML-enabled systems is still weak and dominated by isolated studies.

In order to help addressing these issues, we conducted an international survey with the aim to understand the current industrial RE practices and problems that practitioners face when developing ML-enabled systems. In total, 188 practitioners from 25 countries completely answered the survey. Based on practitioners' responses, we conducted quantitative and qualitative analyses, providing insights into (i) what role is typically in charge of requirements, (ii) how requirements are typically elicited and documented, (iii) which non-functional requirements typically play a major role, (iv) which RE activities are perceived as most difficult, and (v) what RE-related problems do ML practitioners face. We share our findings on the state of practice and problems of RE for ML with the community to help steer future research on the topic.

The remainder of this paper is organized as follows. Section II provides the background and related work. Section III describes the research method. Section IV presents the results. Sections V and VI discuss the results and threats to validity. Finally, Section VII presents our concluding remarks.

## 2  Background and Related Work

ML involves algorithms that analyze data to create models capable of making predictions on new, unseen data [20]. Unlike traditional systems, ML-enabled systems learn from data instead of being programmed with predefined rules. However, poor-quality data can lead to inaccurate results. This supposes a change in the way of designing and developing this type of system. On the other hand, RE constitutes approaches to address challenges that are amplified by the use of ML, e.g., understanding the problem space, aligning interdisciplinary teams, and dealing with stakeholder expectations.

RE and ML have a special connection. According to Kästner [10], an ML model can be seen as a requirements specification based on training data since the data can be seen as a learned description of how the ML model shall behave. In this manner, when developing ML models, we need to identify relevant and representative data, validate models, and balance model-related user expectations (*e.g.*, accuracy versus inference time); just as in RE for traditional systems where we need to identify representative stakeholders, validate specifications with customers, and address conflicting requirements.

Current theoretical SE research has identified many challenges with RE for ML [3,18,19]. Some studies have proposed new methods or adapted existing ones to handle requirements on such systems [9, 26, 27]. While these research contributions are valuable, gathering empirical evidence from the industry is essential to bridge the gap between theory and practice. Collecting practitioners' insights becomes imperative to identify real-world challenges and current practices accurately. Such studies can provide a better understanding of the practical problems that can guide the advancement of new RE for ML techniques and their effective implementation in real-world scenarios. In the following, we present studies conducted within industry settings involving practitioners to understand RE for ML.

Vogelsang and Borg [28] conducted interviews with four data scientists to find out the current practices and what should be done to handle and surpass the challenges regarding requirements. They suggest the need for new RE for ML solutions or at least the adaptation of existing ones. Habibullah *et al.* [8] conducted interviews and a survey to understand how Non-Functional Requirements (NFRs) are perceived among ML practitioners. They identified the degree of importance practitioners place on different NFRs, explored how NFRs are defined and measured, and identified associated challenges.

Recently, Nahar *et al.* [21] identified challenges in building ML-enabled systems through a systematic literature survey aggregating existing studies involving interviews or surveys with practitioners of multiple projects. With respect to RE, they reported challenges related to unrealistic expectations from stakeholders, vagueness in ML problem specifications, and additional requirements such as regulatory constraints. Scharinger *et al.* [22] revealed the worries at Siemens regarding problems that any ML project is susceptible, listing *ML Pitfalls*, such as lack of decision quality baselines and underestimating costs. They believe that RE is the key to avoid this pitfalls and to ripen ML development.

We complement the valuable research discussed above with additional empirical evidence on current practices and problems regarding RE for ML-enabled systems, obtained from an industrial survey on ML-enabled systems.

## 3 Research Method

### 3.1 Goal and Research Questions

The goal of this paper is to characterize the current practices and problems experienced by practitioners in the requirements life cycle stage of ML-enabled system projects. From this goal, we established the following research questions:

- **RQ1. What are the contemporary practices of RE for ML-enabled systems?** This question aims at revealing how practitioners are currently approaching RE for ML, identifying trends, prevalent methods, and the extent to which the industry aligns with established practices. We refined *RQ1* into more detailed questions as follows:
    - RQ1.1 Who is addressing the requirements of ML-enabled system projects?
    - RQ1.2 How are requirements typically elicited in ML-enabled system projects?
    - RQ1.3 How are requirements typically documented in ML-enabled system projects?
    - RQ1.4 Which NFRs do typically play a major role in ML-enabled system projects?
    - RQ1.5 Which activities are considered to be most difficult when defining requirements for ML-enabled system projects?
- **RQ2. What are the main RE-related problems faced by practitioners in ML-enabled system projects?** Identifying these challenges is crucial as it informs the development of strategies to mitigate difficulties, helping to steer future research on the topic in a problem-driven manner. For this research question, we applied open and axial coding procedures to allow the problems to emerge from open-text responses provided by the practitioners.

### 3.2  Survey Design

We designed our survey based on best practices of survey research [30], carefully conducting the following steps:

- **Step 1. Initial Survey Design**. We conducted a literature review on RE for ML [25] and combined our findings with previous results on RE problems [6] and the RE status quo [29] to provide the theoretical foundations for questions and answer options. Therefrom, the initial survey was drafted by software engineering and machine learning researchers of PUC-Rio (Brazil) with experience in R&D projects involving ML-enabled systems.
- **Step 2. Survey Design Review**. The survey was reviewed and adjusted based on online discussions and annotated feedback from software engineering and machine learning researchers of BTH (Sweden). Thereafter, the survey was also reviewed by the other co-authors.
- **Step 3. Pilot Face Validity Evaluation**. This evaluation involves a lightweight review by randomly chosen respondents. It was conducted with 18 Ph.D. students taking a Survey Research Methods course at UCLM (Spain) (taught by the second author). They were asked to provide feedback on the clearness of the questions and to record their response time. This phase resulted in minor adjustments related to usability aspects and unclear wording. The answers were discarded before launching the survey.

- **Step 4. Pilot Content Validity Evaluation**. This evaluation involves subject experts from the target population. Therefore, we selected five experienced data scientists developing ML-enabled systems, asked them to answer the survey, and gathered their feedback. The participants had no difficulties in answering the survey and it took an average of 20 minutes. After this step, the survey was considered ready to be launched.

The final survey started with a consent form describing the purpose of the study and stating that it is conducted anonymously. The remainder was divided into 15 demographic questions (D1 to D15) followed by three specific parts with 17 substantive questions (Q1 to Q17): 7 on the ML life cycle and problems, five on requirements, and five on deployment and monitoring. This paper focuses on the demographics, the ML life cycle problems related to problem understanding and requirements, and the specific questions regarding requirements. The excerpts of the substantive questions related to this paper are shown in Table 1. The survey was implemented using the Unipark Enterprise Feedback Suite.

Table 1: Research questions and survey questions

| RQ | Survey No. | Description | Type |
|---|---|---|---|
| - | … | … | … |
| RQ2 | Q4 | According to your personal experience, please outline the main problems or difficulties (up to three) faced during the Problem Understanding and Requirements ML life cycle stage. | Open |
| - | … | … | … |
| RQ1.1 | Q8 | Who is actively addressing the requirements of ML-enabled system projects in your organization? | Closed (MC) |
| RQ1.2 | Q9 | How were requirements typically elicited in the ML-enabled system projects you participated in? | Closed (MC) |
| RQ1.3 | Q10 | How were requirements typically documented in the ML-enabled system projects you participated in? | Closed (MC) |
| RQ1.4 | Q11 | Which Non-Functional Requirements (NFRs) typically play a major role in terms of criticality in the ML-enabled system projects you participated in? | Closed (MC) |
| RQ1.5 | Q12 | Based on your experience, what activities do you consider most difficult when defining requirements for ML-enabled systems? | Closed (MC) |
| - | … | … | … |

## 3.3 Data Collection

Our target population concerns professionals involved in building ML-enabled systems, including different activities, such as management, design, and development. Therefore, it includes practitioners in positions such as project leaders, requirements engineers, data scientists, and developers. We used convenience sampling, sending the survey link to professionals active in our partner companies, and also distributed it openly on social media. We excluded participants that informed having no experience with ML-enabled system projects. Data collection was open from January 2022 to April 2022. In total, we received responses from 276 professionals, out of which 188 completed all four survey sections. The average time to complete the survey was of 20 minutes. We conservatively considered only the 188 fully completed survey responses.

### 3.4   Data Analysis Procedures

For data analysis purposes, given that all questions were optional, the number of responses varies across the survey questions. Therefore, we explicitly indicate the number of responses when analyzing each question.

Research questions *RQ1.1 - RQ1.5* concern closed questions, so we decided to use inferential statistics to analyze them. Our population has an unknown theoretical distribution (*i.e.*, the distribution of ML-enabled system professionals is unknown). In such cases, resampling methods like bootstrapping, have been reported to be more reliable and accurate than inference statistics from samples [17, 30]. Hence, we use bootstrapping to calculate confidence intervals for our results, similar as done in [29]. In short, bootstrapping involves repeatedly taking samples with replacements and then calculating the statistics based on these samples. For each question, we take the sample of $n$ responses for that question and bootstrap $S$ resamples (with replacements) of the same size $n$. We assume $n$ as the total valid answers of each question [5], and we set 1000 for $S$, which is a value that is reported to allow meaningful statistics [15].

For research question *RQ2*, which seeks to identify the main problems faced by practitioners involved in engineering ML-enabled systems related to problem understanding and requirements, the corresponding survey question is designed to be open text. We conducted a qualitative analysis using open and axial coding procedures from grounded theory [24] to allow the problems to emerge from the open-text responses reflecting the experience of the practitioners. The qualitative coding procedures were conducted by one PhD student and reviewed by her advisor at one site (Brazil) and reviewed independently by three researchers from two additional sites (two from Sweden and one from Turkey).

The questionnaire, the collected data, and the quantitative and qualitative data analysis artifacts, including Python scripts for the bootstrapping statistics and graphs and the peer-reviewed qualitative coding spreadsheets, are available in our open science repository[9].

## 4   Results

### 4.1   Study Population.

Figure 1 summarizes demographic information on the survey participants' countries, roles, and experience with ML-enabled system projects in years. It is pos-sible to observe that the participants came from different parts of the world, representing various roles and experiences. While the figure shows only the ten countries with the most responses, we had respondents from 25 countries. As expected, our convenience sampling strategy influenced the countries, with most responses being from the authors' countries (Brazil, Turkey, Austria, Germany, Italy, and Sweden).

---

[9] https://doi.org/10.5281/zenodo.8248332

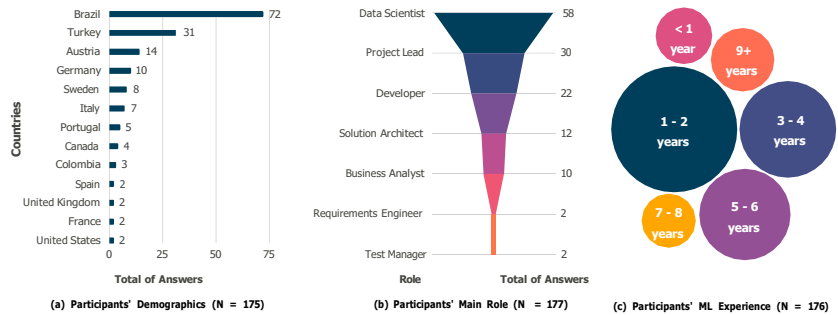

Fig. 1: Demographics: countries, roles, and years of experience.

Regarding employment, 45% of the participants are employed in large companies (2000+ employees), while 55% work in smaller ones of different sizes. It is possible to observe that they are mainly data scientists, followed by project leaders, developers, and solution architects. It is noteworthy that only two participants identified themselves as requirements engineers. Regarding their experience with ML-enabled systems, most of the participants reported having 1 to 2 years of experience. Following closely, another substantial group of participants indicated a higher experience bracket of 3 to 6 years. This distribution highlights a balanced representation of novice and experienced practitioners. Regarding the participants' educational background, 81.38% mentioned having a bachelor's degree in computer science, electrical engineering, information systems, mathematics, or statistics. Moreover, 53.72% held master's degrees, and 22.87% completed Ph.D. programs.

### 4.2 Problem Understanding and Requirements ML Life Cycle Stage

In the survey, based on the nine ML life cycle stages presented by Amershi *et al.* [2] and the CRISP-DM industry-independent process model phases [23], we abstracted seven generic life cycle stages and asked about their perceived relevance and difficulty. The answers, presented in Figure 2, revealed that ML practitioners are extremely worried about requirements. The *Problem Understanding and Requirements* stage is clearly perceived as the most relevant and most complex life cycle stage.

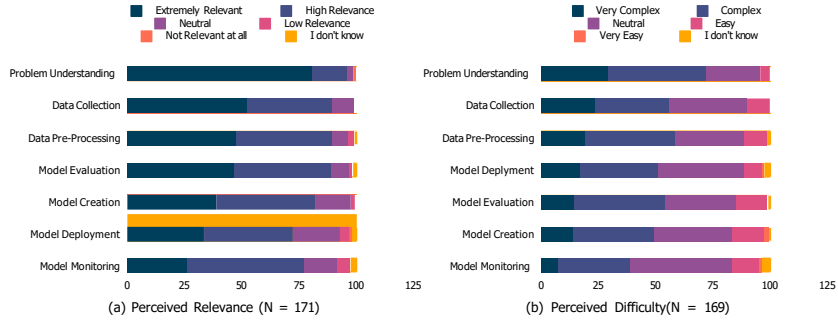

Fig. 2: Perceived relevance and complexity of each ML life cycle stage

### 4.3 Contemporary RE practices for ML-enabled Systems

**[RQ1.1] Who is addressing the requirements of ML-enabled system projects?** The proportion of roles reported to address the requirements of ML-enabled system projects within the bootstrapped samples is shown in Figure 3 together with the 95 % confidence interval. The N in each figure caption is the number of participants that answered this question. We report the proportion P of the participants that checked the corresponding answer and its 95% confidence interval in square brackets.

It is possible to observe that the project lead and data scientists were most associated with requirements in ML-enabled systems with **P = 56.439 [56.17, 56.709]** and **P = 54.71 [54.484, 54.936]**, while Business Analysts (**P = 29.518 [29.288, 29.749]**) and Requirements Engineers (**P = 11.202 [11.061, 11.342]**) had a much lower proportion. Several isolated options were mentioned in the "Others" field (*e.g.*, Product Owner, Machine Learning Engineer, and Tech Lead), altogether summing up 11% and not significantly influencing the overall distribution (**P = 11.021 [10.865, 11.177]**).

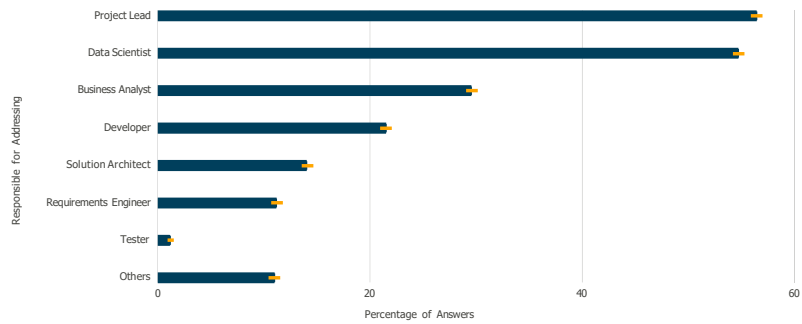

Fig. 3: Roles addressing requirements of ML-enabled systems (N = 170)

**[RQ1.2] How are requirements typically elicited in ML-enabled system projects?** As presented in Figure 4, respondents reported interviews as the most commonly used technique (**P = 55.795 [55.567, 56.022]**), followed (or complemented) by prototyping (**P = 43.953 [43.711, 44.195]**), scenarios (**P = 43.065 [42.834, 43.297]**), workshops (**P = 42.708 [42.483, 42.933]**), and observation **P = 36.838 [36.613, 37.063]**.

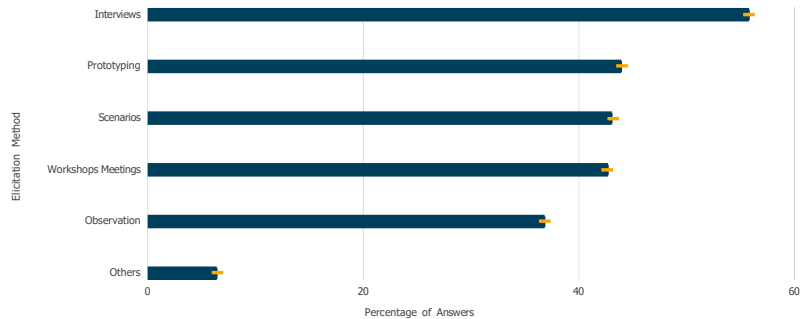

Fig. 4: Requirements elicitation techniques of ML-enabled systems (N = 171)

**[RQ1.3] How are requirements typically documented in the ML-enabled system projects?** Figure 5 shows Notebooks as the most frequently used documentation format with **P = 37.357 [37.149, 37.564]**, followed by User Stories (**P = 36.115 [35.875, 36.356]**), Requirements Lists (**P = 29.712 [29.499, 29.925]**), Prototypes (**P = 23.957 [23.748, 24.166]**), Use Case Models (**P = 21.617 [21.412, 21.822]**), and Data Models (**P = 19.92 [19.724, 20.117]**). Surprisingly, almost 17% mentioned that requirements are not documented at all with **P = 16.955 [16.767, 17.143]**. Several isolated options were mentioned in the "Others" field(*e.g.*, Wiki tools, Google Docs, Jira) with **P = 8.877 [8.744, 9.011]**.

**[RQ1.4] Which Non-Functional Requirements (NFRs) do typically play a major role in terms of criticality in the ML-enabled system projects?** Regarding NFRs (Figure 6), practitioners show a significant concern with some ML-related NFRs, such as data quality (**P = 69.846 [69.616, 70.075]**), model reliability (**P = 42.679 [42.45, 42.907]**), and model explainability (**P = 37.722 [37.493, 37.952]**). Some NFRs regarding the whole system were also considered important, such as system performance (**P = 40.789 [40.573, 41.006]**), and usability (**P = 29.589 [29.36, 29.818]**). A significant amount of participants informed that NFRs were not at all considered within their ML-enabled system projects (**P = 10.617 [10.465, 10.768]**). Furthermore, in the "Others" field (**P = 1.814 [1.745, 1.884]**), a few participants also mentioned that they did not reflect upon NFRs.

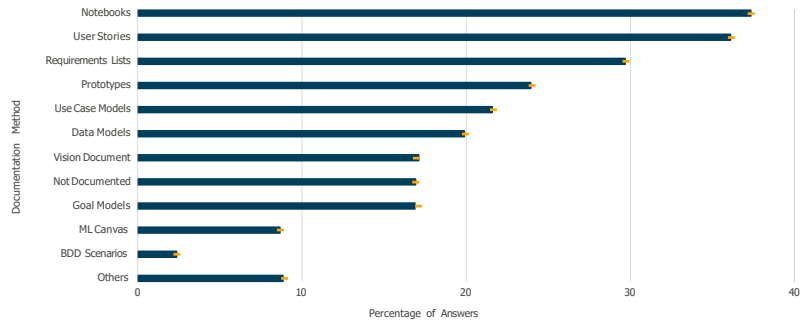

Fig. 5: Requirements documentation of ML-enabled systems (N = 171)

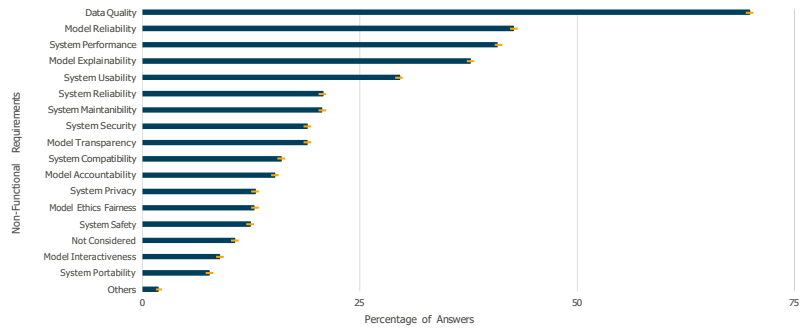

Fig. 6: Critical non-functional requirements of ML-enabled systems (N = 169)

**[RQ1.5] Which activities are considered most difficult when defining requirements for ML-enabled systems?** We provided answer options based on the literature on requirements [29] and requirements for machine learning [25], leaving the "Other" option to allow new activities to be added. As shown in Figure 7, respondents considered that managing customer expectations is the most difficult task (**P = 66.804 [66.575, 67.032]**), followed by aligning requirements with data (**P = 57.306 [57.066, 57.546]**), resolving conflicts (**P = 38.582 [38.341, 38.824]**), managing changing requirements (**P = 35.62 [35.395, 35.846]**), selecting metrics (**P = 33.95 [33.723, 34.176]**), and elicitation and analysis (**P = 29.036 [28.824, 29.248]**).

### 4.4  Main RE-related problems in ML-enabled System Projects

Regarding the main problems faced by the participants during the Problem Understanding and Requirements stage, they emerged from open coding applied to free text answers. Participants could inform up to three problems related to each ML life cycle stage. In total, 262 open-text answers were provided for problems related to problem understanding and requirements.

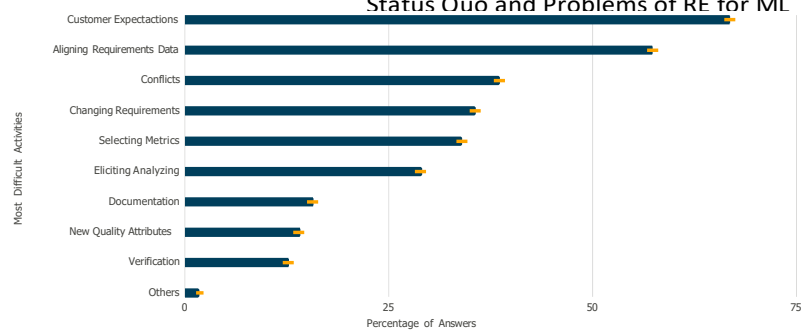

Fig. 7: Most difficult RE activities in ML-enabled systems (N = 171)

We incorporated axial coding procedures to provide an easily understandable overview, relating the emerging codes to categories. We started with the sub-categories *Input*, *Method*, *Organization*, *People*, and *Tools*, as suggested for problems in previous work on defect causal analysis [11]. Based on the data, we merged the *Input* and *People* categories, as it was difficult to separate between the two, given the concise answers provided by the participants. We also renamed the *Tools* category into *Infrastructure* and identified the need to add a new category related to *Data*. It is noteworthy that these categories were identified considering the overall coding for the seven ML life cycle stages, while in this paper, we focus on the problem understanding and requirements stage.

Figure 8 presents an overview of the frequencies of the resulting codes using a probabilistic cause-effect diagram, which was introduced for causal analysis purposes in previous work [12, 13]. While this representation provides a comprehensive overview, the percentages are just frequencies of occurrence of the codes (*i.e.*, the sum of all code frequencies is 100%). Also, the highest frequencies within each category are organized closer to the middle.

It is possible to observe that most of the reported problems are related to the *Input* category, followed by *Method* and *Organization*. Within the *Input* category, the main problems concern difficulties in understanding the problem and the business domain and unclear goals and requirements. In the *Method* category, the prevailing reported problems concern difficulties in managing expectations and establishing effective communication. Finally, in the *Organization* category, the lack of customer or domain expert availability and engagement and the lack of time dedicated to requirements-related activities were mentioned. While we focus our summary on the most frequently mentioned problems, it is noteworthy that the less frequent ones may still be relevant in practice. For instance, computational constraints or a lack of data quality (or availability) can directly affect ML-related possibilities and requirements.

## 5   Discussion

The survey findings reveal an intriguing aspect within ML contexts: the distribution of roles in RE activities. Contrary to conventional expectations, the role of requirements engineers and business analyst appears to be less prominent. Instead, a notable shift is observed, with project leaders and data scientists taking the lead in RE efforts. As the literature suggests that RE can help address problems related to engineering ML-enabled systems, this could point to the fact that software engineering practices are not yet well established within this domain. Nevertheless, the involvement of project leaders and data scientists as key RE contributors could reflect the nature of ML projects, where domain expertise and data-driven insights are pivotal. This shift in responsibilities raises questions about the evolving dynamics of cross-functional collaborations within ML endeavors and prompts further exploration into how such roles influence the shaping of ML-enabled systems.

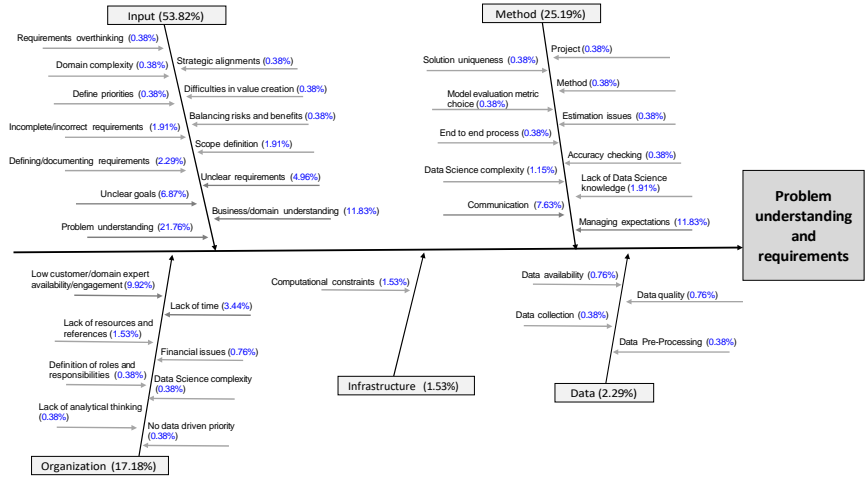

Fig. 8: Main problems faced during problem understanding and requirements

The survey also revealed that practitioners typically use traditional requirements elicitation techniques (interviews, prototyping, scenarios, workshops, and observation). Comparing the results to the elicitation techniques reported for traditional RE [29], an observable difference is that requirements workshops are slightly less commonly used in ML contexts. This could be related to the absence of the requirements engineer, who is typically familiar with conducting such workshops, or to the lack of specific adaptations on such workshop formats for ML-enabled systems.

With respect to requirements documentation, notebooks, which are interactive programming environments that can be used to process data and create ML models, appear as the most used tool for documenting requirements. Again, this could be a symptom of the absence of a requirements engineer and the lack of awareness of RE specification practices and tools. Furthermore, a proportion of almost 17% mentioned that requirements were not documented at all. Given that in conventional contexts problems related to requirements are common causes of overall software project failure [6], this apparent lack of RE-related maturity may also be causing pain in ML contexts. Traditional artifacts, such as user stories, requirements lists, prototypes, and use case models, are also used in the ML context, but significantly less than in the conventional software context [29]. Even specific approaches, such as the ML Canvas, do not relevantly represent a current practice for documenting the requirements of ML-enabled systems.

Regarding NFRs, practitioners express considerable concerns with specific ML-related NFRs, such as data quality, model reliability, and model explainability, while also recognizing the significance of overall system-related NFRs. Nevertheless, more than 10% of practitioners do not even consider NFRs in their ML-enabled system projects. Again, given the potential negative impacts of missing NFRs on software-related projects [6], this can be seen as another indicator of the lack of overall awareness of the importance of RE in the industrial ML-enabled systems engineering context.

The survey also revealed the most difficult activities perceived by practitioners in defining requirements for ML-enabled systems. The difficulties reported by practitioners includes managing customer expectations and aligning requirements with data, highlight the importance of effective communication, a deep understanding of customer needs, and domain and technical expertise to bridge the gap between aspirations and technological feasibility.

Finally, we contributed to the RE-related problems faced by practitioners in ML-enabled system projects. The main issues relate to difficulties in problem and business understanding, managing expectations, and low customer or domain expert availability or engagement. These issues clearly have comparable counterparts in the conventional RE problems [6]. As comparable problems may have comparable solutions, adopting established RE practices (or adaptations of such practices) may help improve ML-enabled system engineering.

## 6   Threats to Validity

We identified some threats while planning, conducting, and analyzing the survey results. Hereafter we list these potential threats, organized by the survey validity types presented in [16].

**Face and Content Validity**. Face and content validity threats include bad instrumentation and inadequate explanation of constructs. To mitigate these threats, we involved several researchers in reviewing and evaluating the questionnaire with respect to the format and formulation of the questions, piloting it with 18 Ph.D. students for face validity and with five experienced data scientists for content validity.

**Criterion Validity**. Threats to criterion validity include not surveying the target population. We clarified the target population in the consent form (before starting the survey). We also considered only complete answers (*i.e.*, answers of participants that answered all four survey sections) and excluded participants that informed having no experience with ML-enabled system projects.

**Construct Validity**. We ground our survey's questions and answer options on theoretical background from previous studies on RE [6, 29] and a literature review on RE for ML [25]. A threat to construct validity is inadequate measurement procedures and unreliable results. To mitigate this threat we follow recommended data collection and analysis procedures [30].

**Reliability**. One aspect of reliability is statistical generalizability. We could not construct a random sample systematically covering different types of professionals involved in developing ML-enabled systems, and there is yet no generalized knowledge about what such a population looks like. Furthermore, as a consequence of convenience sampling, the majority of answers came from Europe and South America. Nevertheless, the experience and background profiles of the subjects are comparable to the profiles of ML teams as shown in Microsoft's study [14]. To deal with the random sampling limitation, we used bootstrapping and only employed confidence intervals, conservatively avoiding null hypothesis testing. Another reliability aspect concerns inter-observer reliability, which we improved by including independent peer review in all our qualitative analysis procedures and making all the data and analyses openly available online.

## 7   Conclusions

Literature suggests that RE can help to tackle challenges in ML-enabled system engineering [25]. Recent literature studies (*e.g.*, [1, 21, 25]) and industrial studies (*e.g.*, [22, 28]) on RE for ML-enabled systems have been important to help to understand the literature focus and industry needs. However, the studies on industrial practices and problems are still isolated and not yet representative.

We complement these studies, aiming at strengthening empirical evidence on current RE practices and problems when engineering ML-enabled systems, with an industrial survey that collected responses from 188 practitioners involved in engineering ML-enabled systems. We applied bootstrapping with confidence intervals for quantitative statistical analysis and open and axial coding for qualitative analysis of RE problems. The results confirmed some of the findings of previous ML-enabled system studies, such as the relevance NFRs related to data quality, model reliability, and explainability [8, 28], and challenges related to customer expectation management and vagueness of requirements specifications [21, 25]. However, we also shed light on some new and intriguing aspects. For instance, the survey revealed that project leaders and data scientists are taking the reins in RE activities for the ML-enabled systems and that interactive Notebooks dominate requirements documentation. With respect to the problems, the main issues relate to difficulties in problem and business understanding, difficulties in managing expectations, unclear requirements, and lack of domain expert availability and engagement.

Overall, when comparing RE practices and problems within ML-enabled systems with conventional RE practices [29] and problems [6], we identified significant variations in the practices but comparable underlying problems. As comparable problems may have comparable solutions, we put forward a need to adapt and disseminate RE-related practices for engineering ML-enabled systems.